\documentclass{aa}
\usepackage{graphics}

\begin{document}

\title{Evolutionary Self-Energy-Loss Effects in Compact Binary
  Systems: Importance of Rapid Rotation and of Equation of State}

 \author{N. K. Spyrou
\and  N. Stergioulas}

\institute{ Astronomy Department,
Aristoteleion University of Thessaloniki,
540.06 Thessaloniki, Macedonia, Greece \\
 {\it email: spyrou@astro.auth.gr, niksterg@astro.auth.gr}
}

\date{Received / Accepted}

\titlerunning{Self-Energy-Loss Effects in Compact Binary Systems}
\maketitle

\maketitle

\begin{abstract}
  The spin-down of a millisecond pulsar in a compact binary system
  leads to self-energy losses, that cause the binary's orbital period
  to increase (the effect being of first post-Newtonian (1PN) order).
  If the pulsar's period-derivative is not exceedingly small, this
  effect can become measurable in long-term high-precision timing
  measurements of the binary's orbital motion. We use rotating
  compressible spheroids to obtain an approximate, explicit expression
  for the orbital period derivative, due to spin-down, valid to 1PN
  order.  This expression can be used to observationally constrain the
  pulsar's moment of inertia, and, combined with other observations,
  the high-density equation of state of compact stars. We apply our
  expression to representative models of millisecond pulsars in binary
  systems and demonstrate the importance of including higher-order
  rotational effects as well as the importance of the choice of
  equation of state, which both have been neglected, so far, in the
  literature.  The computed increase in orbital period is larger, by
  as much as 32\%, when the higher-order rotational effects are
  included, while it can change by more than a factor of two between
  different plausible equations of state.

\end{abstract}

\section{Introduction and Motivation}
\label{introduction}

A large number of known millisecond pulsars are members of binary
systems (see Phinney \& Kulkarni 1994 and Lorimer 1998 for reviews).
Nine new millisecond pulsars in binary systems have recently been
discovered in the globular cluster 47 Tucanae (Camilo et al. 2000),
and future observations are expected to yield an even larger number of
such systems. Long-term timing observations of several relativistic
binary pulsars have allowed the measurement of secular changes in the
orbital period, caused by emission of gravitational radiation, with
very high accuracy (see Kaspi 1999 for a recent review).

Various causes for secular changes in the orbital period of a binary
pulsar have been examined to date. The first and traditional such
study, in the case of a two-point-mass binary, has been based on the
use of the {\it quadrupole formula} for the loss of orbital energy in
the form of gravitational radiation and of the binary's Newtonian
(Keplerian) orbital motion (Peters and Mathews, 1963). The above
lowest-order result has been extended (Spyrou and Papadopoulos 1985)
to the general-relativistic 1PN approximation, with the aid of a
similarly generalized quadrupole formula (Epstein and Wagoner, 1975)
and of the binary's 1PN orbital motion (Spyrou 1981, 1983; see also
Blanchet and Sch{\"a}fer, 1989). Of central importance to the above
calculations is the fact that both members of the binary are treated
as realistic bodies with finite dimensions, internal structure and
internal motions, as opposed to the usual Newtonian point mass of the
standard two-body Kepler problem. More specifically, the members are
described in terms of the so-called {\it inertial} (or {\it
  gravitational}) mass, namely, their total mass-energy, and the
corresponding position and velocity three-vectors. These parameters
are valid to 1PN approximation and, initially, they were proposed by
Spyrou (1977a,b;1978, see also Caporali and Spyrou, 1981 for an
inclusion of finite-size effects in a relativistic parametrized
description of the realistic binary). As a consequence, the binary's
total inertial mass manifests itself in the 1PN description of its
orbital motion and, therefore, any of the members's characteristics of
internal structure and motions can in principle affect its orbital
motion and hence orbital period.  Especially in the case of the binary
pulsar B1913+16, it was proved (Spyrou and Papadopoulos, 1985) that,
based on the then available observational data (Weisberg and Taylor
1984), the relative 1PN correction to the orbital-period shortening,
deduced from the above mentioned generalized quadrupole formula, was
about five orders of magnitude smaller than the relative observational
uncertainty.

Another cause for secular changes in the orbital period
of a binary or binary-participating compact star, are evolutionary
changes in the compact star's total self-energy (Spyrou 1985, 1987;
see also Spyrou 1988, 1999 for the importance, as a cause of
orbital-period changes, of the accretion-induced changes of
binary-participating pulsar's period of axial rotation and radius).
Especially in the case of the binary pulsar B1913+16, it was claimed
(Spyrou 1987) that an accuracy improvement in the next decimal place
of the then observed values (Weisberg and Taylor, 1984) of the orbital
period's rate of change and of the associated uncertainty, would permit
orbital-period shortening due to self-energy losses to be
distinguished from that due to gravitational-radiation losses.

In this paper, we will focus on the secular spin-down of a pulsar (due
e.g. to magnetic braking), that can cause a decrease in its rotational
kinetic energy and thus, a corresponding change in the pulsar's
gravitational mass. By conservation of the orbital angular momentum,
this leads to an increase in the orbital period of the binary (Jeans
1924, 1928, Smarr \& Blandford 1976, Damour \& Taylor 1991). If the
pulsar's period is only a few milliseconds and its spin-down rate is
of the order of $10^{-19}$s/s (which is not uncommon for millisecond
pulsars), then the induced secular increase in the orbital period may
be measurable with current technology.  Previous studies of this
effect have used only lowest-order approximations for the self-energy
change of the pulsar, to obtain order-of-magnitude estimates of the
effect (see e.g. Smarr \& Blandford 1976, Will 1981, Damour \& Taylor
1991). For example, the star was considered to be spherical, i.e.
higher order rotational corrections to the shape of the star were
neglected, and no particular equation of state was chosen. Here, we
present a detailed description of mass-energy loss from a rapidly
rotating pulsar, deriving an expression for the orbital period change
that can be used to accurately constrain the pulsar's moment of
inertia, from long-term timing measurements of the orbital period. Our
treatment relies on the description of rapidly rotating neutron stars
as compressible ellipsoids (Lai, Rasio \& Shapiro, 1993, hereafter
LRS). We apply our results to typical millisecond pulsars in binary
systems and derive constraints on their parameters, for the effect to
be observable.  Moreover, our results are derived consistently to
first post-Newtonian order. The relation between mass-energy
loss and orbital period increase, originally derived by Jeans (1924),
assumed Newtonian gravity and isotropic mass-energy loss. The
inclusion of 1PN terms in the description of the binary's orbit,
changes the orbital period at 2PN order. Thus, Jean's relation is
valid to 1PN order. Finally, we show that, including spin-orbit coupling
in the description of the binary's orbit, changes the above relation
by a term that can be neglected for typical binary systems. 

The outline of the paper is as follows: In the next Section 2, we derive
the self-energy loss of a rapidly rotating compact star, due to evolutionary
spin-down, while in Section  3, we derive the corresponding orbital period 
change. In Sections 4 and 5 we present lowest-order estimates and exact
numerical results for the orbital period change and we conclude with a
discussion of our results in Section 6.

\section{Self-energy loss due to spin-down}
\label{mass-energy}

We will derive a relation between the spin-down rate of a pulsar and
the resulting mass-energy loss, valid for rapidly rotating stars. The
structure of rotating stars is approximated by {\it compressible
  spheroids} (introduced by LRS). These are approximate equilibrium
configurations, the compressible generalization of {\it Maclaurin
  spheroids} (Chandrasekhar 1969), constructed using a variational
method. The equation of state (EOS) has the usual zero-temperature
polytropic form
\begin{equation}
  \label{eq:EOS}
  P=K \rho^{1+1/n},
\end{equation}
where $K$ is the polytropic constant and $n$ is the polytropic index.
A polytropic index of $n=0.5$ corresponds to a relatively
stiff EOS, while a polytropic index of $n=1$ corresponds to a
relatively soft EOS. The rigidly-rotating, oblate spheroid has major
axis $a_1$, minor axis $a_3<a_1$ and eccentricity
\begin{equation}
  \label{eq:eccentricity}
  e=\sqrt{1-(a_3/a_1)^2}.
\end{equation}
If $\bar m$ is the baryonic mass of the spheroid and $R_0$ the radius of a
{\it nonrotating} (spherical) star of the same baryonic mass, then the
gravitational binding energy $W$ is  
\begin{equation}
  \label{eq:W}
  W=-\frac{3}{5-n}\frac{G \bar m^2}{R_0}g(e)\left[g(e)\left( 1-2\frac{T}{|W|} 
         \right) \right]^{n/(3-n)},
\end{equation}
where,
\begin{equation}
  \label{eq:g}
  g(e)=(1-e^2)^{1/6}\frac{\sin^{-1}e}{e}.
\end{equation}
In (\ref{eq:W}), the ratio of rotational kinetic energy $T$ to $|W|$ depends
only on the eccentricity $e$ and is given by
\begin{equation}
  \label{eq:T/W}
  \frac{T}{|W|}=\frac{3}{2e^2}\left(1-\frac{e(1-e^2)^{1/2}}{\sin^{-1}e} \right) 
                 -1.
\end{equation}
The internal energy $U$ of the spheroid is assumed to satisfy the
virial relation
\begin{equation}
  \frac{3}{n}U+W+2T=0,
\end{equation}
and the total self-energy ${\cal E} = U+W+T$ is given by
\begin{equation}
  \label{eq:E}
  {\cal E}=\frac{3-n}{3}W \left(1- \frac{3-2n}{3-n} \frac{T}{|W|} \right).
\end{equation}
The moment of inertia is
\begin{equation}
  \label{eq-I}
  I=\frac{2}{5}\kappa_n \bar m R_0^2(1-e^2)^{-1/3}\left[g(e)\left( 1-2\frac{T}{|W|} 
          \right) \right]^{-\frac{2n}{3-n}},
\end{equation}
where $\kappa_n$ is a constant that depends on the equation of state (see
LRS), with typical values $\kappa_n=(1;0.82;0.65)$ for $n=(0;0.5;1)$. Then,
the rotational period of the star is defined through
\begin{equation}
  \label{eq:P}
  P=2 \pi \sqrt{I/2T}.
\end{equation}
Thus, through (\ref{eq:W}), (\ref{eq:T/W}) and (\ref{eq-I}), the total
self-energy and the rotational period can be expressed entirely in 
terms of $e$, $\bar m$, $R_0$ and $\kappa_n$.

As the star spins down due to magnetic braking, it follows a
quasi-equilibrium sequence of {\it constant baryonic mass} (we do not
consider significant mass-loss from the pulsar). Along this sequence,
the total self-energy and rotational period, for a given equation of
state, are then functions of the eccentricity only. In order to derive
useful relations, we employ a series expansion of the structure of
rotating compressible spheroids, around the nonrotating configuration,
in powers of the eccentricity $e$, and retain terms of order at most
six in $e$, a quite accurate, indeed, approximation for rotational
periods of a few milliseconds.  The total self-energy and angular
velocity $\Omega=2\pi/P$ are then written as
\begin{eqnarray}
  {\cal E} &=& - \frac{3-n}{5-n} \frac{G \bar m^2}{R_0} \biggl[ 1  -  
            \frac{2}{5(3-n)}e^2  
            -  \frac{9}{35(3-n)}e^4 \nonumber \\
           && - \frac{2(2050-1383n+227n^2)e^6}{2625(3-n)^3} 
            + O_8 \biggr] ,\label{E}  \\
  \Omega &=& e \sqrt{\frac{2G\bar m}{|5-n|\kappa_nR_0^3}} 
           \biggl[ 1-\frac{3(11n-5)}{70(3-n)}e^2  \nonumber \\
          &&  -\frac{9(25+982n-416n^2)}{9800(3-n)^2}e^4 \nonumber
 \label{P}
\end{eqnarray}
\begin{eqnarray}
            -\frac{(3490875+12039717n-11829047n^2+\frac{7669405}{3}n^3)}
                 {7546000(3-n)^3} e^6 \nonumber
\end{eqnarray}
\begin{equation}
          + O_8 \biggr].
\end{equation}
Inverting the series expansion of $\Omega$, one obtains the following 
expansion for the eccentricity:
\begin{eqnarray}
  e = \frac{\tilde \Omega}{\sqrt{2}} \Biggl[ 1 &+& \frac{3(11n-5)}{140(3-n)} 
        \tilde \Omega^2 +
        \frac{397n^2+414n+225}{2800(3-n)^2} 
        \tilde \Omega^4  \nonumber \\  
         &+& O_6 \Biggr],\label{e}
\end{eqnarray}
where 
\begin{equation}
  \tilde \Omega = \Omega \sqrt{\frac{|5-n|\kappa R_0^3}{G \bar m}},
\end{equation}
is a dimensionless form of the angular velocity. Taking time derivatives
(denoted by an overdot), one can express the mass-energy loss $\dot {\cal E}$
in terms of the spin-down rate $\dot P$:
\begin{eqnarray}
  \dot {\cal E} = - 4 \pi^2 I_0 \frac{\dot P}{P^3} \Biggl[ &1& + 
     \frac{6 \pi^2(5+n)}{5(3-n)}\frac{1}{\tilde P^2} \nonumber \\
       &+& \frac{4\pi^4(275+204n+n^2)}{35(3-n)^2}\frac{1}{\tilde P^4} + O_6 
      \Biggr], \label{Edot}
\end{eqnarray}
where $\tilde P = 2\pi/ \tilde \Omega$ is a dimensionless form of the
rotational period and $I_0=\frac{2}{5}\kappa_n\bar m R_0^2$ is the moment
of inertia of a nonrotating star of baryonic mass $\bar m$ and radius
$R_0$. With this definition, $R_0$ remains constant during spin-down
(when changes in the baryonic mass are neglected), while the
equatorial and polar radii can both change due to the changing
eccentricity.

\section{Orbital Period Change}

The {\it gravitational} mass $m$ of a compact star (entering the 
gravitational equations of motion) can be decomposed (Spyrou 1977a) as 
\begin{equation}
  m= \bar m+\frac{\cal E}{c^2},
\label{m}
\end{equation}
where $\bar m$ is the baryonic rest mass and $\cal E$ is
the Newtonian self-energy, i.e
\begin{equation}
  {\cal E}=W+U+T
\end{equation}
We consider the 1PN motion of a binary system, consisting of two compact
objects of gravitational masses $m_{\rm p}$ (primary) and $m_{\rm c}$ (companion). 
The orbital period of the binary system of total mass $M=m_{\rm c}+m_{\rm p}$
and orbital energy $E=\mu E_0$ (where $E_0$ is the orbital energy per
unit reduced gravitational mass) is (Blanchet \& Sch{\"a}fer, 1989)
\begin{equation}
 \label{Porb}
  P_{\rm b}= \frac{2\pi GM \mu^{3/2}}{(-2E)^{3/2}}\left[1+\frac{1}{4}
            \left(\frac{\mu}{M}-15 \right)\frac{E}{\mu c^2} \right],
\end{equation}
and $\mu=m_{\rm c}m_{\rm p}/(m_{\rm c}+m_{\rm p})$ is the reduced
gravitational mass of the system. If the primary loses energy
isotropically and the orbital angular momentum is conserved, then it has been
shown by Jeans (1924,1928), assuming mass-energy equivalence and
Keplerian orbital motion, that the induced change in the binary's
orbital period is
\begin{equation}
\label{Pd}
 \frac{\dot P_{\rm b}}{P_{\rm b}}= -2\frac{ \dot m_{\rm p}}{M}.
\end{equation}
It is easy to show that under our assumptions, the 1PN term on the
r.h.s. of (\ref{Porb}) does not contribute to 1PN order in (\ref{Pd}).

We notice that, the pulsar's spin can be included in the equations
describing the orbital motion, following, e.g. the formalism of
Gergely, Perj{\'e}s \& Vas{\'u}th (1998). In the Appendix, we show, that
including the spin-orbit coupling, the total change in the binary's
orbital period is
\begin{equation}
\label{Pd_spin}
 \frac{\dot P_{\rm b}^{\rm S}}{P_{\rm b}}= - 2\frac{ \dot m_{\rm p}}{M}
 - \frac{3G(2+\eta) \cos \psi }{2c^2a^{3/2}\sqrt{GM(1-\epsilon^2)}} \dot S,  
\end{equation}
where $\eta=m_{\rm c}/m_{\rm p}$, $\psi$ is the angle substended by the
orbital angular momentum $\vec L$ and the pulsar's spin $\vec S$, $a$
and $\epsilon$ are the semi-major axis and eccentricity of the orbit and $S$
is the magnitude of the spin. In (\ref{Pd_spin}), only the dominant, 
to 1PN order, spin-terms are included. In the Appendix, we also show
that, for typical millisecond binary pulsars, the spin-term in the
r.h.s.  of (\ref{Pd_spin}) is of order $(R_0/a)^{3/2}$ smaller than the
first term. Since we will not be concerned with binaries that
are near coalescence, in the remainder of this paper, we will neglect 
the spin-term in (\ref{Pd_spin}) and use only equation (\ref{Pd}).

Assuming that the spin-down is mainly due to magnetic braking and, thus, 
neglecting changes in the baryonic mass of the primary, we find
\begin{equation}
  \dot m_{\rm p}= \frac{\dot {\cal E}}{c^2},
\end{equation}
and through (\ref{Edot}), (\ref{m}) and (\ref{Pd}), we derive our final
expression for the orbital change due to spin-down
\begin{eqnarray}
   \frac{\dot P^{\rm S}_{\rm b}}{P_{\rm b}} = \frac{8 \pi^2 I_0}{c^2M} \frac{\dot P}{P^3} 
      & \Biggl[& 1 + 
     \frac{6 \pi^2(5+n)}{5(3-n)}\frac{1}{\tilde P^2} \nonumber \\
      &+& \frac{4\pi^4(275+204n+n^2)}{35(3-n)^2}\frac{1}{\tilde P^4} + O_6 
      \Biggr], \label{EdS}
\end{eqnarray}
This expression depends only on the spin and spin-down rate of the
primary, the total mass $M$ of the binary system and the moment of
inertia of a {\it nonrotating} star of same baryonic mass as the
primary. To lowest order in the eccentricity, our result agrees with
slow-rotation expressions derived previously in the literature (see
Smarr \& Blandford 1976, Will 1981, Damour \& Taylor
1991)\footnote{Note that, the first two references contain a missprint
that was corrected in the third reference.}, and extends those results
to rapidly rotating stars. As we will show in the remainder of this
paper, the higher order rotational corrections that we introduce in
(\ref{EdS}) are essential, if one wants to arrive at quantitative
constraints on the structure of compact objects, using observationally
derived data of the above effect.

\section{Lowest Order Estimates}
 
Before applying (\ref{EdS}) to specific binary systems, it is instructive
to use the lowest-order term 
\begin{equation}
   \left(\frac{\dot P^{\rm S}_{\rm b}}{P_{\rm b}} \right)_0 \simeq
 \frac{8 \pi^2 I_0}{c^2M} \frac{\dot P}{P^3}, 
      \label{EdS_0}
\end{equation}
for obtaining a qualitative conclusion about the possible types
of binary systems, in which the orbital change due to spin-down can be
observationally important.  For the effect to be measurable, it must
be larger than achievable timing precision. For example, the precision
with which the orbital decay $\dot P_{\rm orb}/P_{\rm orb}$ (due to
quadrupole gravitational radiation emission) of the binary pulsar
B1913+16 has been measured is $2.3 \times 10^{-19}{\rm s}^{-1}$ (Taylor 1992, 1993;
Kaspi 1999).

Assuming a radius of $R=12$km, a spin period of $1.6$ms and assuming
that $m_{\rm c} \ll m_{\rm p}$, we see that the present effect exceeds
the above precision, if $\dot P>3\times 10^{-20} {\rm s/s} $. In the next
section, we will show that including the higher order corrections in
rotation, relaxes this requirement. Typical values of $\dot P$ for
pulsars with periods of a few milliseconds are $10^{-21} {\rm s/s}
<\dot P <10^{-19} {\rm s/s} $ (Phinney \& Kulkarni 1994, Lorimer 1998).
Thus, we conclude that in binary systems in which the primary is a
rapidly rotating millisecond pulsar with periods of a few milliseconds
only and spin-down rate larger than a few times $10^{-20} {\rm s/s}$,
the orbital change due to the pulsar's spin-down becomes, in
principle, observable. In contrast, the secular increase in the
orbital period of the relativistic binary pulsar B1913+16 (due to the
pulsar's spin-down) is several orders of magnitude smaller than the
secular decrease caused by emission of gravitational radiation,
because of the pulsar's relatively large spin period of 59ms (see
Spyrou 1987 for a related discussion - an upper limit to
self-energy-loss effects has still not been reached by current
observational accuracy). Hence, in the case of B1913+16 and similar
relativistic binary systems, the influence of self-energy losses on
the orbital period are still beyond reach of experimental
verification.

\section{Results for Typical Binary Pulsars}

We now apply equation (\ref{EdS}) to different possible millisecond pulsars in
binary systems. We use the accuracy in measuring changes in the
orbital period of PSR B1913+16, as a measure of the observability of the
changes induced by spin-down. Thus, we define the ratio
\begin{eqnarray}
  \Lambda = \frac{\dot P^{\rm S}_{\rm b} / P_{\rm b}}{ 2.3 \times 10^{-19}{\rm s}^{-1}}.
\end{eqnarray}
We note that, the measured value of $\dot P /P$ for millisecond pulsars
is normally corrected by a galactic acceleration term. In the case
of PSR B1913+16, the uncertainty in estimating the galactic acceleration
term in $\dot P /P$ is $1.9 \times 10^{-19} {\rm s}^{-1}$ (Damour and Taylor 1991),
which is slithgly less than the presently achieved accuracy in measuring
$\dot P /P$ for this pulsar..

In order to illustrate the importance of including the higher-order
rotational terms, we define
\begin{equation}
  \Delta=\frac{\dot P_b^{\rm S}-(\dot P_b^{\rm S})_0}{\dot P_b^{\rm S}},
\end{equation}
as being the relative difference between the result (\ref{EdS}) (which
includes terms up to $O_4$) and the lowest order estimate
(\ref{EdS_0}). We focus on typical millisecond pulsars of mass
$m_p=1.41 M_\odot$ using two different equations of state that span the
expected range of possible realistic EOSs: a relatively stiff EOS,
with polytropic index $n=0.5$ and a relatively soft EOS, with
polytropic index $n=1.0$. For the stiff EOS, we choose the nonrotating
radius to be $R_0=12$km, while for the soft EOS, the nonrotating
radius is chosen to be $R_0=10$km. In this way, the corresponding
nonrotating moment of inertia $I_0$ is, in both cases, in agreement
with the relativistic moment of inertia computed by Ravenhall \&
Pethick (1994) for various equations of state. Rapidly rotating neutron
stars may have mass larger than $1.4 M_\odot$, if they are spun-up by accretion.
However, the neutron star's mass enters in the r.h.s. of eq. (\ref{EdS}) only
through the ratio $I_0/M \sim R_0^2$, which is much more sensitive to the radius of
the star than to its mass. Thus, it is sufficient, in the present context,
to consider only one representative value for the mass.

\begin{table}
\begin{center}
\begin{tabular}{*{4}{c}}
\multicolumn{4}{c}{}\\
\multicolumn{4}{c}{\bf Stiff EOS ($n=0.5$)}\\
\multicolumn{4}{c}{}\\
\hline
\hline
\multicolumn{4}{c}{}\\
$\dot P$ & $\dot P^{\rm S}_{\rm orb}/P_{\rm orb}$  & $\Delta $ & 
$\Lambda$   \\[0.5ex]
 s/s     & $({\rm s}^{-1})$   & $(\%)$  &    \\[0.5ex]
\hline
\multicolumn{4}{c}{$P$=1.6 ms}\\
\hline
\\[0.2ex]
$1\times 10^{-18}$  &  $1.3\times 10^{-17}$    & 32  &  56.4      \\[0.5ex]
$5\times 10^{-19}$  &  $6.5\times 10^{-18}$    & 32  &  28.2      \\[0.5ex]
$1\times 10^{-19}$  &  $1.3\times 10^{-18}$    & 32  &   5.6      \\[0.5ex]
$5\times 10^{-20}$  &  $6.5\times 10^{-19}$    & 32  &   2.8      \\[0.5ex]
$1\times 10^{-20}$  &  $1.3\times 10^{-19}$    & 32  &   0.6      \\[0.5ex]
\multicolumn{4}{c}{}\\
\hline
\multicolumn{4}{c}{$P$=3 ms}\\
\hline
\\[0.2ex]
$5\times 10^{-18}$  &  $7.4\times 10^{-18}$    & 10  &  32.4      \\[0.5ex]
$1\times 10^{-18}$  &  $1.5\times 10^{-18}$    & 10  &  6.5      \\[0.5ex]
$5\times 10^{-19}$  &  $7.4\times 10^{-19}$    & 10  &  3.2      \\[0.5ex]
$1\times 10^{-19}$  &  $1.5\times 10^{-19}$    & 10  &  0.7       \\[0.5ex]
$5\times 10^{-20}$  &  $7.4\times 10^{-20}$    & 10  &  0.3      \\[0.5ex]
\end{tabular}
\vspace{3mm}
\caption{Relative rate of orbital period change (second column), 
  corresponding to different values of the pulsar's period derivative
  $\dot P$. Two different cases for the pulsar's period are
  considered.  The quantity $\Delta$ measures the significance of the
  higher-order rotational corrections, while $\Lambda$ is the ratio of the
  second column to the accuracy achieved in measuring orbital period
  changes in the binary pulsar B1913+16.  The pulsar's mass is $m_{\rm
    p}=1.41 M_\odot$, while the mass of the companion is $m_{\rm c}=0.2
  M_\odot$. Results are for a stiff equation of state, of polytropic index
  $n=0.5$.}
\label{tab1}
\end{center}
\end{table}

\begin{table}
\begin{center}
\begin{tabular}{*{4}{c}}
\multicolumn{4}{c}{}\\
\multicolumn{4}{c}{ \bf Soft EOS ($n=1.0$)}\\
\multicolumn{4}{c}{}\\
\hline
\hline
\multicolumn{4}{c}{}\\
$\dot P$ & $\dot P^{\rm S}_{\rm orb}/P_{\rm orb}$  & $\Delta $ & 
$\Lambda$   \\[0.5ex]
 s/s     & $({\rm s}^{-1})$   & $(\%)$  &    \\[0.5ex]
\hline
\multicolumn{4}{c}{$P$=1.6 ms}\\
\hline
\\[0.2ex]
$1\times 10^{-18}$  &  $6.0\times 10^{-18}$    & 19  &  26.1      \\[0.5ex]
$5\times 10^{-19}$  &  $3.0\times 10^{-18}$    & 19  &  13.1      \\[0.5ex]
$1\times 10^{-19}$  &  $6.0\times 10^{-19}$    & 19  &   2.6      \\[0.5ex]
$5\times 10^{-20}$  &  $3.0\times 10^{-19}$    & 19  &   1.3      \\[0.5ex]
$1\times 10^{-20}$  &  $6.0\times 10^{-20}$    & 19  &   0.3      \\[0.5ex]
\multicolumn{4}{c}{}\\
\hline
\multicolumn{4}{c}{$P$=3 ms}\\
\hline
\\[0.2ex]
$5\times 10^{-18}$  &  $3.9\times 10^{-18}$    & 6  &  17.0      \\[0.5ex]
$1\times 10^{-18}$  &  $7.8\times 10^{-19}$    & 6  &  3.4      \\[0.5ex]
$5\times 10^{-19}$  &  $3.9\times 10^{-19}$    & 6  &  1.7      \\[0.5ex]
$1\times 10^{-19}$  &  $7.8\times 10^{-20}$    & 6  &  0.3       \\[0.5ex]
$5\times 10^{-20}$  &  $3.9\times 10^{-20}$    & 6  &  0.2      \\[0.5ex]
\end{tabular}
\vspace{3mm}
\caption{Same as Table 1, but for a soft equation of state of polytropic index $n=1.0$.}
\label{tab2}
\end{center}
\end{table}

The mass of the
companion is fixed at $m_{\rm c}=0.2M_\odot$, which is a typical value for
white-dwarf companions of millisecond pulsars in binary systems (our
results are weakly dependent on the mass of the companion, if the
companion is a low-mass white dwarf). We do not consider the case of
binary systems with two neutron stars, as, in that case, changes in
the orbital period are dominated by quadrupole gravitational radiation
losses (see Spyrou \& Kokkotas 1994, for an account of self-energy loss
effects on the gravitational radiation from neutron star binaries)

Table 1 displays our results for the stiff EOS. For $P=1.6$ms
(eccentricity $e=0.517$), $\Lambda=56.4$ for $\dot P=10^{-18}$ s/s, while
$\Lambda=5.6$ for $\dot P=10^{-19}$ s/s is also interesting for
observations. Notice the large relative difference of $\Delta = 32\% $,
which demonstrates the importance of the higher-order rotational
effects. For $P=3$ ms (eccentricity $e=0.273$), $\Lambda=32.4$ for $\dot P=5
\times 10^{-18}$ s/s and $\Lambda=3.2$ for $\dot P=5\times 10^{-19}$ s/s, while, for
$\dot P< 10^{-19}$ s/s, the magnitude of the effect is less than the
presently achieved observational accuracy. For the larger rotational
period, the effect of higher-order rotational terms is reduced to $\Delta =
10\% $.

Table 2, displays the corresponding results for the soft EOS. For
$P=1.6$ms (eccentricity $e=0.332$), $\Lambda=26.1$ for $\dot P=10^{-18}$
s/s, while $\Lambda=2.6$ for $\dot P=10^{-19}$ s/s, roughly a factor of two
smaller than for the stiff EOS.  The rotational effects are reduced to
$\Delta = 19\% $, For $P=3$ ms (eccentricity $e=0.175$), $\Lambda=17.0$ for $\dot
P=5\times 10^{-18}$ s/s and $\Lambda=1.7$ for $\dot P=5\times 10^{-19}$ s/s.  The
rotational effects are now only $\Delta = 6\% $.
 
It is evident that the results are sensitive to the chosen equation of
state. The present uncertainty in the high-density equation of state
of neutron star matter (which could be as stiff as a $n=0.5$ polytrope
or as soft as a $n=1.0$ polytrope) introduces, in the computed orbital
period change, an unknown factor, which could be as large as a factor
of two. A measurement of the orbital change due to the pulsar's
spin-down and a knowledge (or guess) of it's baryonic mass could yield
a value for (or constrain) the moment of inertia $I_0$ of nonrotating
neutron stars of same mass $m_{\rm p}$.  Combined with a measurement
(or educated guess) of the pulsar's mass, this information can then
constrain the very high-density equation of state of compact stars.

\section{Discussion}                                              

We have extended previous, approximate, calculations of the orbital
period change in binary pulsars (due to the pulsar's spin-down), by
including higher-order rotational effects and the influence of the
choice of the high-density equation of state. We have shown that the
above higher-order terms, as well as the choice of the equation of
state are significant, if quantitative constraints are to be drawn for
future observations of this effect. We also showed, that the inclusion
of the pulsar's spin in the description of the binary's orbit, does
not contribute significantly to the orbital period change. Our
treatment relies on the description of rotating stars as compressible
perfect fluid spheroids (ignoring the possible influence of a solid
crust), in which one can express both the change in spin-period and
the change in radius, during spin-down, through the change in the
eccentricity of the spheroid. Binary pulsars, that are interesting for
the above effect to be measurable, must be rapidly rotating, with a
rotational period of only a few ms and, in addition, with a period
derivative equal to or larger than a few times $10^{-20} {\rm s/s}$.
As Phinney \& Kulkarni (1994) point out, a millisecond pulsar like 
B1937+21 (which has $P=1.56$ms and $\dot P=10^{-19}$s/s), if it were in
a binary system, would yield a measurable orbital period change. Using
our results, i.e. equation (\ref{EdS}), one could then derive
quantitative constraints on the pulsar's moment of inertia. More
accurate constraints could be derived, using numerical models of
rapidly rotating neutron stars that satisfy the general-relativistic
field equations, for various realistic equations of state. Such a
computation, however, will make sense only after observational
measurements of the predicted effect will start becoming available.
The increasing number of millisecond pulsars discovered in binary
systems (Camilo et al. 2000) is a positive sign that this could happen
in the near future.

\section*{Acknowledgements}                                              

We thank Bernard F. Schutz, Gerhard Sch{\"a}fer, Kostas Kokkotas and
Theocharis Apostolatos for helpful comments and suggestions and
John H. Seiradakis for a careful reading of the manuscript.


\section*{Appendix}

We have shown that (\ref{Pd}) depends only on the Newtonian orbital
characteristics (as the 1PN orbital terms contribute at 2PN order).
Thus, the pulsar's spin can be included in the description of the
binary's orbit, following the formalism of Gergely, Perj{\'e}s \& Vas{\'u}th
(1998), who consider the correction to the Newtonian orbit, induced by
the spin.  We define $\eta=m_{\rm c}/m_{\rm p}$, $\psi $ the angle
substended by the orbital angular momentum $\vec L$ and the pulsar's
spin $\vec S$, $a$ and $\epsilon$ are the semi-major axis and eccentricity of
the pulsar's orbit and $L$, $S$ the magnitudes of its orbital angular
momentum and spin. Then, the binary's period $P_b$, orbital energy $E$
and semi-major axis $a$, are given by
\begin{eqnarray}
  P_b&=&2 \pi \frac{GM \mu^3}{(-2 \mu E)^{3/2}}, \\
  E &=& -\frac{GM \mu}{2a} \left[1+\frac{GS \cos \psi  (2+\eta)}
                       {c^2 a^{3/2}\sqrt{GM(1-\epsilon^2)}} \right],
\end{eqnarray}
and
\begin{equation}
  a = -\frac{GM\mu}{2E} \left[ 1- \frac{2ES \cos \psi }{c^2ML}(2+\eta) \right]
\end{equation}

In the case that the spin is zero and the star loses mass-energy 
isotropically (e.g. due to thermal emission), it has been shown by Jeans
(1924, 1928) that the product $A=aM$ remains constant during
the orbit's evolution (which leads to equation (\ref{Pd}) in the text).
We treat the inclusion of spin in the description of the orbital motion
as a first-order perturbation about the orbital motion with zero spin.
If so, we can write
\begin{eqnarray}
 E &=& {\cal E}_0 (1+\delta \tilde E), \\ 
 A &=& A_0(1+\delta \tilde A), 
\end{eqnarray}
where
\begin{equation}
  \delta \tilde E = \frac{GS \cos \psi  (2+\eta)}{c^2 a^{3/2}\sqrt{GM(1-\epsilon^2)}},
\end{equation}
\begin{equation}
  \delta \tilde A = -\frac{2ES \cos \psi }{c^2ML}(2+\eta) -\delta \tilde E, 
\end{equation}
and ${\cal E}_0$, $A_0$ are the the values of $E$ and $A$ when spin-orbit
coupling is ignored. Then,
\begin{equation}
  \frac{\dot P_b^{\rm S}}{P_b} = -2\frac{ \dot m_{\rm p}}{M}+\frac{3}{2} \left(
    \dot{\delta \tilde A} - \dot{\delta \tilde E} \right).
\end{equation}
Keeping only first-order terms in the spin (and keeping only 1PN terms),
we find 
\begin{equation}
\label{Pd_spin2}
 \frac{\dot P_{\rm b}^{\rm S}}{P_{\rm b}}= - 2\frac{ \dot m_{\rm p}}{M}
 - \frac{3G(2+\eta) \cos \psi }{2c^2a^{3/2}\sqrt{GM(1-\epsilon^2)}} \dot S,  
\end{equation}
In order to arrive at an order-of-magnitude estimate for the spin-term in
(\ref{Pd_spin2}), we substitute the lowest-order (in spin-period) 
expressions derived in the main text (using $S=I\Omega$), and find 
\begin{eqnarray}
\label{Pd_spin3}
 \frac{\dot P_{\rm b}^{\rm S}}{P_{\rm b}}&=&\frac{8 \pi^2 I_0}{c^2M} \frac{\dot P}{P^3}  
  \Biggl[ 1+ \nonumber \\
  && \frac{3}{8} \frac{\cos \psi  (2+\eta)}{e\sqrt{1-\epsilon^2}}\sqrt{2|5-n|\kappa_n} 
    \Biggl(\frac{R_0}{a} \Biggl)^{3/2} \Biggr].
\end{eqnarray}
Since the spin-term is of order $(R_0/a)^{3/2}$ smaller than the isotropic
term in (\ref{Pd_spin3}), it can be neglected for typical binary pulsar
systems considered in the present paper.

\end{document}